\documentclass[12pt,preprint]{aastex}

\begin{document}

\title{Natural Coronagraphic Observations of the Eclipsing T Tauri
System KH 15D: Evidence for Accretion and Bipolar Outflow in a 
WTTS\footnotemark[1]}

\author{Catrina M. Hamilton\altaffilmark{2} and William Herbst}
\affil{Astronomy Department, Wesleyan University, Middletown, CT
06459}
\email{catrina@astro.wesleyan.edu,bill@astro.wesleyan.edu}

\author{Reinhard Mundt and Coryn A. L. Bailer-Jones}
\affil{Max-Planck-Institut f\"{u}r Astronomie, K\"{o}nigstuhl 17,
D-69117 Heidelberg, Germany}
\email{mundt@mpia-hd.mpg.de,calj@mpia-hd.mpg.de}

\and

\author{Christopher M. Johns-Krull}
\affil{Physics and Astronomy Department, Rice University, Houston, TX 77005}
\email{cmj@rice.edu}

\footnotetext[1]{Based on Ultraviolet-Visual Echelle Spectrograph observations
collected at the European Southern Observatory's Very Large Telescope via
Director's Discretionary Time, within the observing program P267.C-5736.}

\altaffiltext{2}{Physics Department, Wesleyan University, Middletown, CT
06459}

\begin{abstract}

We present high resolution (R $\sim$ 44,000) UVES
spectra of the eclipsing
pre-main sequence star KH 15D covering the wavelength
range  4780 to 6810 {\AA} obtained at three phases: out
of eclipse, near minimum light and during egress. The system
evidently acts like a natural coronagraph, enhancing the contrast
relative to the continuum of hydrogen and forbidden emission lines
during eclipse. At maximum light the  H$\alpha$ equivalent width
was $\sim$2 {\AA} and the profile showed broad wings and a deep
central absorption. During egress the equivalent width was much
higher ($\sim$70 {\AA}) and the broad wings, which extend to
$\pm$ 300 km/s, were prominent. During eclipse totality the
equivalent width was less than during egress ($\sim$40 {\AA}) and the
high velocity wings were much weaker. H$\beta$ showed a somewhat
different behavior, revealing only the blue-shifted portion of the high
velocity component during eclipse and egress.
[OI] $\lambda\lambda$6300, 6363 lines are easily seen both
out of eclipse and when the photosphere is obscured and exhibit little
or no flux variation with eclipse phase.  Our interpretation is that KH 15D,
although clearly a weak-line T Tauri star by the usual criteria, is still
accreting matter from a circumstellar disk, and has a well-collimated
bipolar jet.  As the knife-edge of the occulting matter passes across
the close stellar environment it is evidently revealing structure in the
magnetosphere of this pre-main sequence star with unprecedented spatial
resolution. We also show that there is only a small, perhaps marginally
significant, change in the velocity of the K7 star between the
maximum light and egress phases probed here.

\end{abstract}

\keywords{stars:individual (KH 15D) --- accretion --- line:profiles}

\section{Introduction}

KH 15D is a unique system in which a pre-main sequence star
is periodically occulted by extended, non-luminous matter, presumably
part of a circumstellar disk \citep{Herbst02}. Every 48.36 days
the star fades over  2-3 days by $\sim$3.5 mag and
remains near minimum light for $\sim$20 days. The eclipse duration has been
increasing with time, by $\sim$1-2 days/yr. \citet{Ham01}
obtained low resolution spectra in and out of eclipse and concluded that
the star was a K7 weak-line T Tauri star (WTTS).
The WTTS classification, based on the equivalent width (EW) of the
H$\alpha$ line out of eclipse, is supported by the
lack of any IR\footnote{H-K = 0.14, K-L = 0.03 out-of-eclipse, K. Haisch,
private communication.} or UV excess emission and the absence of substantial
photometric variability outside of eclipse.  Based on its membership in
NGC 2264 \citep{Sung97}, the estimated mass of the K7 star is 0.5-1 M$_\odot$
and its age is 2-4 Myr.

In an attempt to learn more about this unique system we obtained 
high resolution spectra during the
eclipse of December 2001. We hoped to determine whether KH
15D is a radial velocity variable, whether there was evidence for
additional light in the system beyond that of the K7 star, and
whether the occultation had any effect on the spectrum. In
fact, we found dramatic changes in the line profiles of
H$\alpha$ and H$\beta$, as well as weak forbidden emission lines
that become much more visible during eclipse. Evidently, the KH 15D
system behaves like a ``natural coronagraph", allowing us to see details of
its close circumstellar environment during eclipse. Our evidence 
suggests
that this WTTS is actively accreting gas
and driving a bipolar outflow, although probably not at the rate
of a typical CTTS. This calls into question the common practice of
associating WTTS characteristics with the absence of an accretion disk.

\section{Observations and the Absorption Spectrum}

High resolution echelle spectra of KH 15D were obtained on the nights of
UT 2001 Nov. 29, when it was in its bright state just
prior to eclipse, on UT 2001 Dec. 14, just past mid-eclipse,
and again on UT 2001 Dec. 20 during egress (see Figure 1).
These data were collected with the
Ultraviolet-Visual Echelle Spectrograph (UVES) on the European Southern
Observatory's Very Large Telescope (VLT), at Mount Paranal, Chile.
The wavelength range is $\sim$4780 to 6810 \AA.  A 50 {\AA} gap
centered on 5800 {\AA} is present due to use of the red arm
which employs a mosaic of two 4096 x 2048 CCDs \citep{D00}.
With a 1\arcsec\ slit, the spectral resolution is $\sim$44,000.  The spectra
presented here were reduced via the UVES pipeline. Information
regarding the reduction process can be found at
http://www.eso.org/instruments/uves/.
As a check on this, the spectra were also reduced in the manner
described by \citet{Valenti94}.  Both procedures make use of a sky subtraction
algorithm.  No significant differences were found between the data reduced by
these techniques and the UVES pipeline reduction is adopted here.

The UVES spectra confirm that, out of eclipse, the primary light
source in the KH 15D system is a K7 WTTS.
The 29 Nov. spectrum was visually compared in detail to 61 Cyg B, a
typical K7 V, to look for any evidence of a second source of light
in the system. No additional features were found. This spectrum was also
compared to a rotationally broadened, synthetic K7 V template to determine its
$v\thinspace sin(i)$, which we estimate to be $<$ 5 km/s.
The EW of the LiI 6707 feature is 0.401 {\AA} $\pm$ 0.001,
consistent with the value of 0.47 {\AA} $\pm$ 0.05 measured on a low
resolution spectrum by \citet{Ham01}.

The out-of-eclipse spectrum was cross-correlated against HD
55999, a standard observed on the same nights,
to determine the radial velocity of KH 15D.
This was done with the IRAF\footnote{Image Reduction and
Analysis Facility, written and supported by the IRAF programming group at
the National Optical Astronomy Observatories (NOAO) in Tucson, Arizona.
NOAO is operated by the Association of Universities for Research in
Astronomy (AURA), Inc. under cooperative agreement with the National
Science Foundation.} task FXCOR.  Since the spectra were obtained on two
CCDs covering different wavelength regions, a cross-correlation was
performed on each, avoiding the emission line regions of H$\alpha$,
and H$\beta$.  A heliocentric radial velocity of +9.0 km/s $\pm$ 0.2,
was found by this procedure for KH 15D on 29 Nov 2001.

The egress spectrum was cross-correlated against the out-of-eclipse
spectrum in the same manner to look for any radial velocity
variation between 29 Nov. and 20 Dec.  A difference in heliocentric
radial velocity of 3.3  $\pm$ 0.6 km/s was found.  The radial velocity
of the star on 20 Dec.
was, therefore, +12.3 km/s $\pm$ 0.6.  Whether this detection
of radial velocity variation means that the K7 star is a spectroscopic
binary remains to be seen. Photospheric line profile variations are expected in
a partially eclipsed star and the likely importance of scattered radiation
near minimum light further complicates the interpretation. Clearly a
more extensive radial velocity study of the system is needed and is
underway (J. Johnson \&  G. Marcy, private communication).

\section{The Emission Line Spectrum}

In Figures 2-4, we show the H$\alpha$, H$\beta$, and [OI] 6300 emission-line
profiles for KH 15D.  Each spectrum has been flux calibrated to a relative
scale using an R magnitude for the date, as given in Table 1. The quoted
uncertainties on the magnitudes reflect a small degree of
non-simultaneity in the spectral and photometric data as well as the
need to transform from I (where the data were more numerous) to R.
Since spectral and photometric data were taken within an hour of each
other, except at maximum light (when the brightness is nearly
constant with time) and since the color variation is quite small
($\sim$0.1 mag in R-I over the full brightness range), the
uncertainty in R is only $\sim$0.1 mag. Julian dates for each
observation, measured I magnitudes, derived R and V magnitudes,
measured EWs, and derived H$\alpha$ and [OI] fluxes are
listed in Table 1.  The EW measurements refer to the
entire profile, both red- and blue-shifted components combined.
On Figs. 2 and 3, the arrows indicate where HI nebular emission
could have affected the profile, whereas on Fig. 4 the arrow indicates
where [OI] $\lambda$6300 from the night sky may not have been removed
completely.

Figure 2 shows that dramatic changes occurred in the H$\alpha$ emission line
profile of KH 15D during eclipse and egress.  Out of eclipse, the star appears
to be a WTTS with an EW(H$\alpha$) $\sim$2 {\AA}, but with a 
double-peaked profile, having a central absorption
and faint, but clearly detectable broad wings. A comparison with the
H$\alpha$ profiles of  other WTTSs \citep{Hartmann82, Mundt83, FB87, 
Edwards94, Reipurth96} shows
that only 3 out of 19 such stars exhibited double-peaked profiles similar to
KH 15D, with UX Tau A being the most similar (see \citet{Reipurth96}).
Most WTTSs show narrow, single-peaked emission lines. It is also
interesting that the central absorption feature appears to extend below the
stellar continuum. This is unusual for any TTS, weak or classical.
During eclipse and egress, the ``natural coronagraph" effect is clear.
Near mid-eclipse, the EW of H$\alpha$ grows to
$\sim$40 {\AA}, while the relative flux drops by $\sim$50\%.
During egress the EW rises to $\sim$70 {\AA}, as the flux increases,
exceeding even its out-of-eclipse value.  In all cases, a
double-peaked profile is observed. The emission line profile during
mid-eclipse extends to $\pm$ 200 km/s with significantly less flux in
the extended wings, as compared to the profile during egress, which
extends to $\pm$ 300 km/s.

Figure 3 shows the corresponding H$\beta$ line profiles.  Again,
dramatic EW and line profile changes are evident.  Although the
out-of-eclipse H$\beta$ line profile is heavily affected by the
underlying stellar absorption spectrum, it is similar
to H$\alpha$. In the egress spectrum, however,
the emission occurs primarily on the blue side with a wing extending
to -300 km/s. The asymmetry in this line is much more striking than 
H$\alpha$.  
During mid-eclipse, although
little H$\beta$ flux is present,
the shape of the line profile appears similar to egress.
The weak features visible in this profile at - 25 km/s are probably due to an
improper background subtraction of the H$\beta$ emission from NGC 2264.

Figure 4 shows the emission-line profiles for the [OI] 6300 {\AA} line. 
The [OI] line, which is weak, but clearly discernable at maximum light,
becomes prominent at mid-eclipse and during egress.
The line flux seems to be about the
same out of eclipse and during egress, although slightly higher at mid-eclipse.
We caution against any extreme interpretation of this measurement.
The profile during mid-eclipse was disturbed by an improper background
subtraction, and we feel that our errors are most likely underestimated.
However, this behavior would indicate that none of the [OI] emitting zone
suffers variable occultation at the phases of our observations.
This is fully expected given the spatial extent of the (bipolar)
forbidden line emitting regions in CTTSs.  The peak of the high velocity
component of the [OI] emission in typical CTTSs originates at about 30 AU
from the star \citep{Hirth97}.

The EW of the [OI] $\lambda$6300 line in the out-of-eclipse
spectrum is about 0.17 {\AA}, which is quite large for a WTTS.  Only
one of the ten WTTSs in Table 3 of \citet{Hartigan95} has a
detectable [OI] $\lambda$6300 line (with an EW = 0.5 {\AA}). The
others have upper limits of about 0.06 {\AA}.
Most CTTSs in that table, on the other hand, have EWs of 0.5 - 3
{\AA}. The profiles obtained during mid-eclipse and egress
suggest that we have two emission peaks, at about -20 and +18 km/s with
emission wings extending -60 to +50 km/s, respectively.  These profiles are
quite different from the [OI] $\lambda$6300 profile seen in most
strong-emission CTTSs (with strong veiling), which often have a high-velocity
component at -100 to -150 km/s (resulting from the jet) and a low-velocity
component at about -20 km/s \citep{Hartigan95, Hirth97}.  In addition, the
profile is quite different from that of a CTTS with small veiling and IR excess
(see the bottom profile in Figure 11 of \citet{Hartigan95}).  These latter
profiles are usually single-peaked and unshifted in velocity.  Although
the profile shape for KH 15D is not fully clear due to imperfect subtraction
of the emission of [OI] at 6300 {\AA} in the night sky, it is most
likely double-peaked.
Such a profile can be most easily explained by a bipolar jet moving nearly
perpendicular to the line of sight, quite consistent with our expectation
that the disk associated with KH 15D is viewed nearly edge on, resulting
in a very small radial velocity separation between the two jets.
The profile catalog of \citet{Hartigan95} contains several examples of
bipolar jet sources (RW Aur, AS 353A, DD Tau), where the two jet
components are clearly separated due to favorable inclination angles.

\section{Discussion}

It is important to keep in mind that we do not yet
know the geometry of the KH 15D system. In particular,
it is not certain whether the occultation proceeds along a line
perpendicular to or parallel to the rotation axis of the K7 star (or
neither) or to what extent the rotation axis, the presumed stellar
magnetic axis and the orbital plane are aligned.
It is not even known whether, relative to the system's center of
mass, it is the K7 star or the occulting matter which is primarily in
motion. Detailed modeling of the system must obviously await
clarification of these basic issues. However, a preliminary
qualitative interpretation of the emission line variations is
possible and provided in this section.

We believe that the observed spectral variations of KH 15D can be understood
qualitatively in terms of a weakly accreting, ``scaled down" classical T Tauri
star (see e.g., \citet{Muz01}) whose photosphere and magnetosphere
are periodically occulted by a relatively sharp knife-edge (as
described in \citet{Herbst02}, their Fig. 5). We expect such a star
to have a bipolar jet region, which we assume, by analogy with the
CTTS,  is revealed by the forbidden emission line radiation. Assuming
a jet velocity of 200 km/s (which is the average jet velocity for
CTTSs derived by \citet{Hirth94}) and adopting a radial velocity of
$\pm$ 20 km/s for the [OI] $\lambda$6300 peaks, we derive an inclination
angle for the jet to the line of sight of 84$\degr$. This is consistent with a
general picture for the system in which we view the K7 star close to
the orbital plane of its circumstellar (or circumbinary) disk and the
jets emerge roughly perpendicular to the disk plane, as for the star
associated with HH 30 and other examples imaged by the Hubble Space
Telescope \citep{Ray96}. The absence of any significant variation in
the profiles or flux of the forbidden line radiation during eclipse
is consistent with the expectation, based on the CTTS analogy, that
it arises at distances of tens of AU's from the star, beyond the region
variably occulted.

The behavior of the hydrogen lines is complex because some
components do arise close to the star and, therefore, suffer variable
occultation effects along with the photosphere. Since we expect that
the H$\alpha$ line has a much higher optical depth than the H$\beta$
line, it is obvious that any emission line region will be more
extended in H$\alpha$ than H$\beta$.  If we assume that the
magnetic axis of the K7 star is tilted toward us at
$\sim$ 5-10$\degr$, which is a reasonable assumption supported
by the [OI] line profiles, at mid-eclipse, one can qualitatively understand the
H$\alpha$ line profile as resulting from low velocity material in the outer,
more extended H$\alpha$ emission region while the star and the H$\beta$
line forming region are obscured by the occulting disk material.  This
would also explain why there is almost no H$\beta$ flux during mid-eclipse.
During egress, we expect most of the H$\alpha$ emission line region, as well
as some of the H$\beta$ emission line region, which is closer to the star,
to be visible.  The H$\alpha$ line profile during egress has a low velocity
emission peak with a central absorption feature similar to what is seen in
the out-of-eclipse profile.  Additionally, two ``shoulders'' appear along the
profile at about $\pm$ 150 km/s extending out to $\pm$ 300 km/s.  These
``shoulders'' could be due to material rotating in the outer parts of the
magnetosphere.  However, given that the $v\thinspace sin(i)$ is measured to be
$<$ 5 km/s, and that the magnetosphere only extends out to
about 5-6 stellar radii, the maximum rotational velocity is about
5-6 $v\thinspace sin(i)$ or 25-30 km/s, making it difficult to
explain the H$\alpha$ emission-line profile with a rotating magnetopshere.

A more attractive hypothesis is that the high velocity
``shoulders'' on the H$\alpha$ line, so prominent in the egress
spectrum, arise from material falling along magnetic accretion
columns.  This interpretation can also qualitatively account for the H$\beta$
emission line profile during egress.  Adopting values for the mass and
radius of the K7 star from Table 1 of \citet{Ham01}, a free-fall
velocity of about 380 km/s can be associated with material at the surface of
the star.  Since H$\beta$ is produced much closer to the star, the
blue-shifted wing extending to nearly -350 km/s could be representative of
material accreting along magnetic field lines near the pole.  The asymmetry
seen in the H$\beta$ emission line profile is most likely due to the fact
that the star is slightly inclined toward our line of sight. The
reappearance of both a blue and red wing to the H$\alpha$ line during
the early part of egress, when most of the stellar photosphere is
still occulted, shows that both red-shifted and blue-shifted gas is
present along the same line of sight towards the small portion of the
photosphere and magnetosphere that is being uncovered first. This
also supports an accretion interpretation as opposed, say, to a
rotation interpretation for the high velocities.

\section{Conclusions}

It appears that KH 15D, although clearly a WTTS by the usual criteria
of weak H$\alpha$ emission, absence of UV or IR excess emission, and
relative photometric stability, is still undergoing active accretion
and driving a bipolar outflow. It provides a cautionary example
against assuming that  all stars with WTTS characteristics no longer
have accretion disks. The unique geometry of this system, in which a
relatively sharp-edged occulting mask crosses the photosphere has
created a ``natural cornagraph" which enhances the visibility of the
star's magnetosphere during eclipse. The occultation also evidently
crosses the inner portion of the star's magnetosphere, where high
velocity gas motions arise, probably from magnetically channeled
accretion. Synoptic studies of this star may ultimately allow us to
reconstruct aspects of the structure of its magnetosphere with
spatial resolution that will be unobtainable in other objects for
decades to come.

\acknowledgements

We thank U. Bastian, M. Ibraghimov,  J. Johnson, G. Marcy, F. Vrba,
and J. Aufdenberg for helpful conversations and useful data on this star.
W.H. acknowledges support from NASA through its Origins of Solar Systems 
program.

\clearpage

\begin{deluxetable}{|c|c|c|c|c|c|c|c|}
\tabletypesize{\scriptsize}
\tablecaption{Journal of Observations and Measurements. \label{tbl-1}}
\tablehead{
\colhead{JD (2452...)} & \colhead{V(mag)} & \colhead{R(mag)} &
\colhead{I(mag)} & \colhead{EW$_{(H\alpha)}$ ({\AA})} &
\colhead{EW$_{(OI)}$ ({\AA})} & \colhead{Flux$_{(H\alpha)}$ (x
10$^{-6}$)\tablenotemark{a}} & \colhead{Flux$_{[OI]}$ (x
10$^{-6}$)\tablenotemark{a}} }

\startdata

242.7296 & 16.10 $\pm$ 0.02 & 15.27 $\pm$ 0.02 & 14.49 $\pm$ 0.02 &
2.11 $\pm$ 0.08 & 0.22 $\pm$ 0.04 & 1.6 $\pm$ 0.06 & 0.17 $\pm$
0.03\\
257.7840 & 19.6 $\pm$ 0.1 & 18.9 $\pm$ 0.1 & 18.2 $\pm$ 0.1 & 39.3
$\pm$ 1.5 & 8.4\tablenotemark{b} $\pm$ 0.2 & 1.1 $\pm$ 0.04 & 0.24 $\pm$ 0.02\\
263.6871 & 19.1 $\pm$ 0.1 & 18.3 $\pm$ 0.1 & 17.6 $\pm$ 0.1 & 70.8
$\pm$ 5.1 & 3.1 $\pm$ 0.2 & 3.4 $\pm$ 0.24 & 0.15 $\pm$ 0.02\\

\enddata

\tablenotetext{a}{Calculated as EW$_{(H\alpha)}$ x 10$^{-0.4*R}$}
\tablenotetext{b}{This EW was measured with the spike removed.}

\end{deluxetable}

\clearpage

\begin{figure}
\plotone{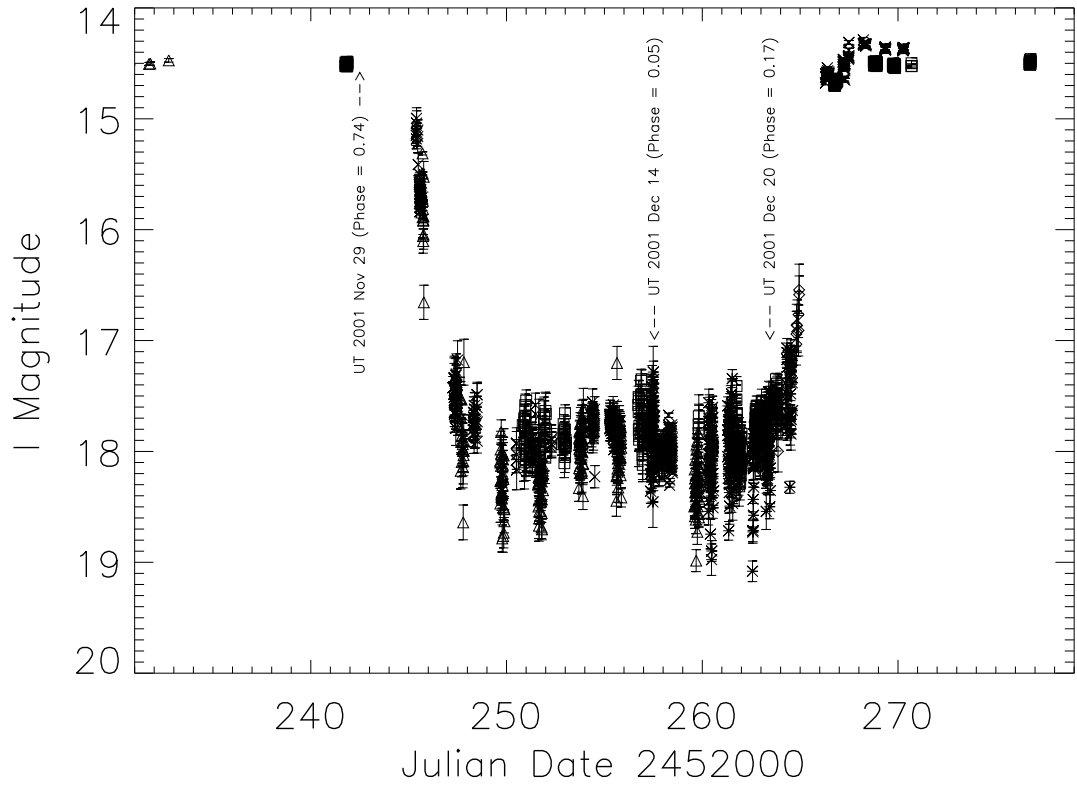}
\figcaption{The December 2001 eclipse of KH 15D; photometry from
Herbst et al. (2002).  Arrows indicate epochs at which the UVES
spectra were obtained.\label{Fig. 1}}
\end{figure}

\begin{figure}
\plotone{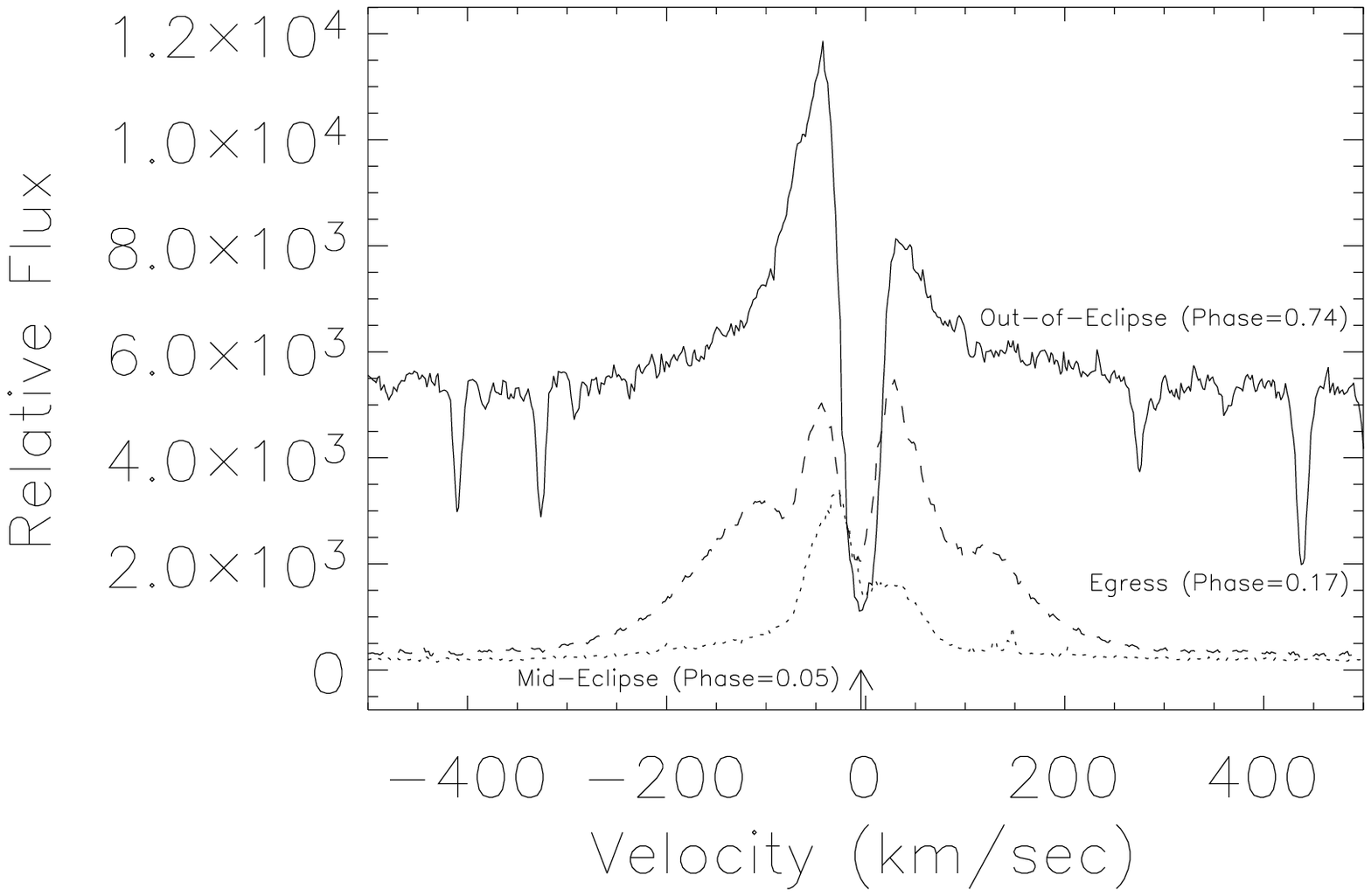}
\figcaption{H$\alpha$ profiles of KH 15D obtained with the VLT and UVES
during the December 2001 eclipse.  The arrow indicates where the nebular
emission of H$\alpha$ is seen in the background.  The velocities are shown
in the reference frame of the star.\label{Fig. 2}}
\end{figure}

\begin{figure}
\plotone{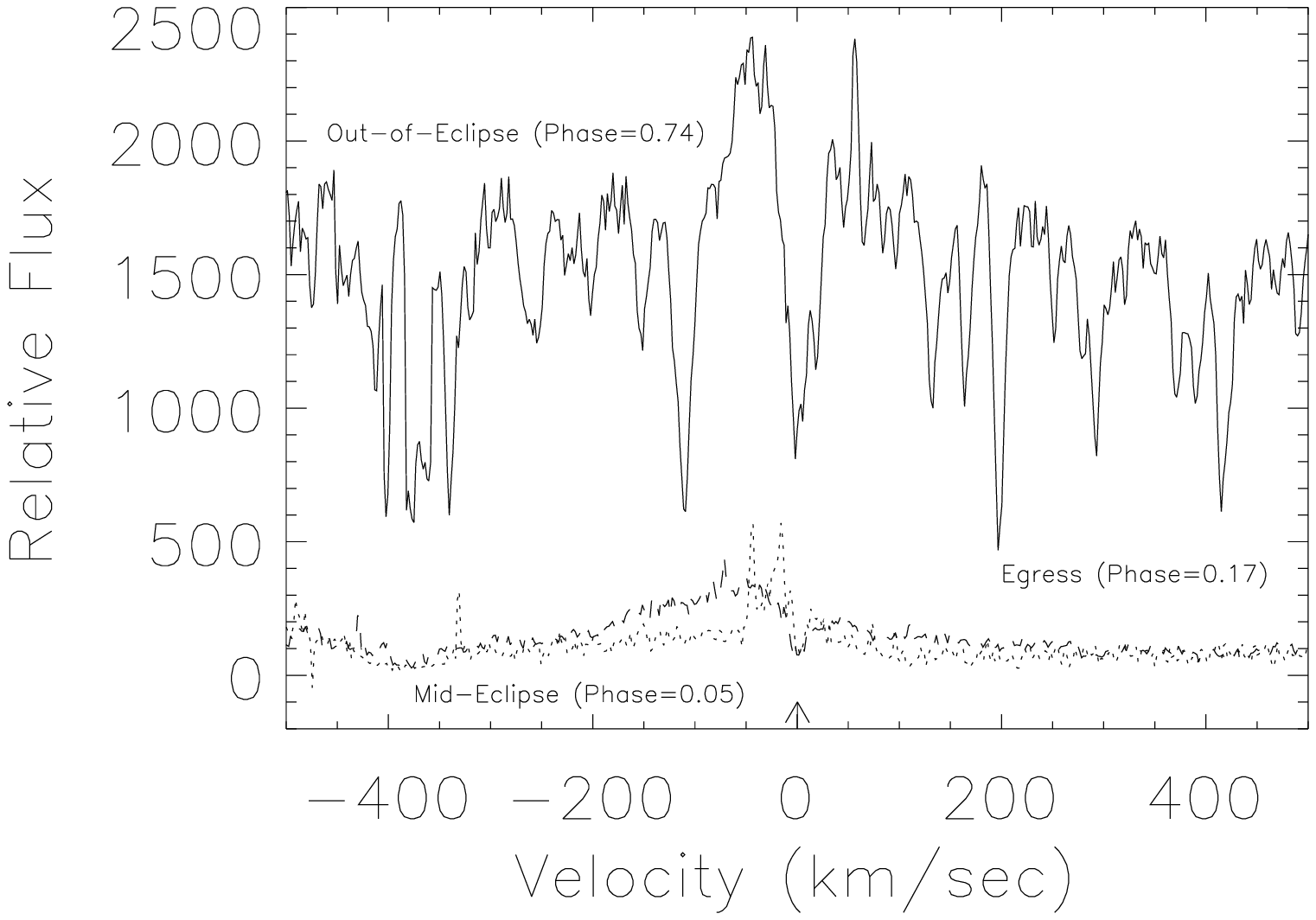}
\figcaption{H$\beta$ profiles of KH 15D obtained with the VLT and UVES
during the December 2001 eclipse.  The arrow indicates where the nebular
emission of H$\beta$ is seen in the background.  The velocities are shown
in the reference frame of the star. \label{Fig. 3}}
\end{figure}

\begin{figure}
\plotone{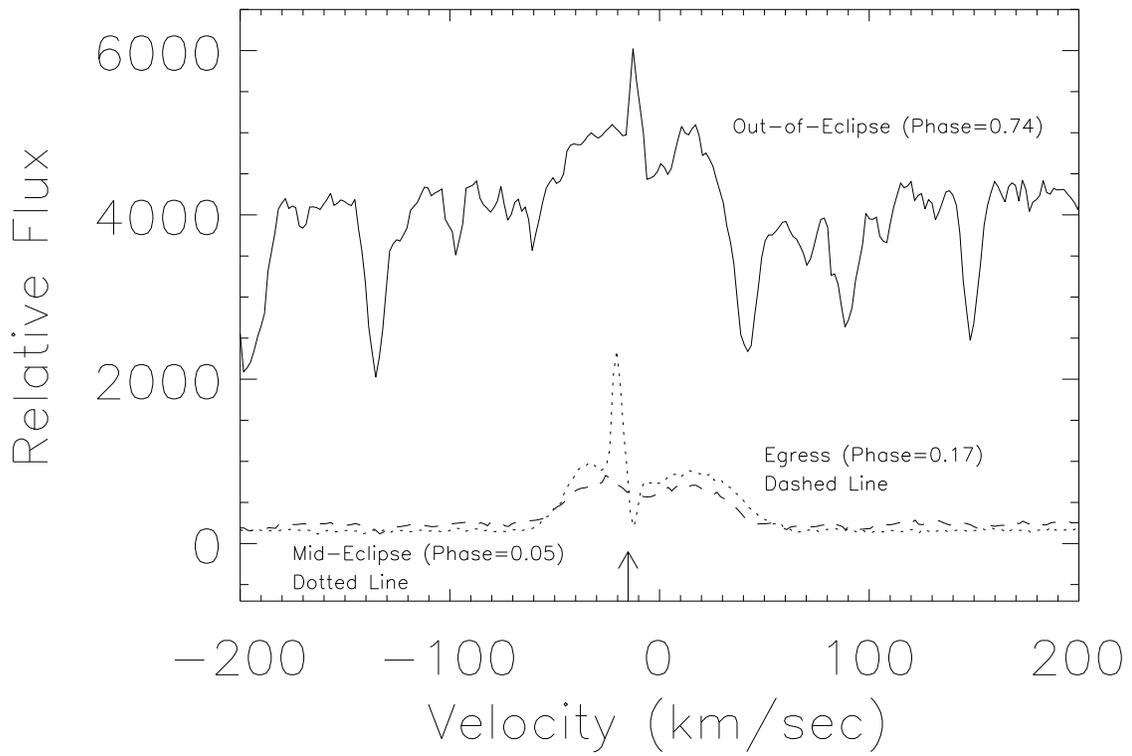}
\figcaption{[OI] 6300 {\AA} profiles of KH 15D obtained with the VLT and UVES
during the December 2001 eclipse.  The arrow indicates where the
night sky emission of [OI] has not been removed completely.  The
velocities are shown
in the reference frame of the star. \label{Fig. 4}}
\end{figure}

\end{document}